\documentclass[11pt]{revtex4}
\usepackage{amssymb,epsf}
\usepackage{latexsym}

\begin{document}

\title{Thermodynamics of charged topological dilaton black holes}
\author{Ahmad Sheykhi \footnote{sheykhi@mail.uk.ac.ir}}
\address{Department of Physics, Shahid Bahonar University, P.O. Box 76175-132, Kerman,
Iran}

\begin{abstract}
A class of $(n+1)$-dimensional $(n\geq3)$ topological black hole
solutions in Einstein-Maxwell-dilaton theory with Liouville-type
potentials for the dilaton field is presented. In these
spacetimes, black hole horizon and cosmological horizon can be an
$(n-1)$-dimensional positive, zero or negative constant curvature
hypersurface. Because of the presence of the dilaton field, these
topological black holes are neither asymptotically flat nor
(anti)-de Sitter. We calculate the charge, mass, temperature,
entropy and electric potential of these solutions. We also analyze
thermodynamics of these topological black holes and disclose the
effect of the dilaton field on the thermal stability of the
solutions.

\end{abstract}
\maketitle
\section{Introduction}
It is generally believed that in asymptotically flat spacetime,
the topology of the event horizon of a stationary black hole in
four dimensions, is uniquely determined to be the two-sphere $S^2$
\cite{Haw1,Haw2}. The ``topological censorship theorem" of
Friedmann, Schleich and Witt is another indication of the
impossibility of non spherical horizons \cite{FSW1,FSW2}. This
theorem states that in a globally hyperbolic, asymptotically flat
spacetime satisfying the null energy condition, any two causal
curves extending from past to future null infinity are homotopic.
A black hole with toroidal surface topology would provides a
possible violation of topological censorship theorem, as a light
ray from past infinity linking with the hole of the torus and then
back to future infinity would not be deformable to a light ray
traveling from past to future outside the black hole. Thus the
hole must quickly close up, before a light ray can pass through
\cite{Jacob}. Therefore, general relativity does not allow an
observer to probe the topology of spacetime, and any topological
structure collapses too quickly to allow light to traverse it.

However, when the asymptotic flatness and the four dimensional
spacetime are given up, there are no fundamental reasons to forbid
the existence of static or stationary black holes with nontrivial
topologies. For instance, for five-dimensional asymptotically flat
stationary black holes, in addition to the known $S^3$ topology of
event horizons, stationary black hole solutions with event
horizons of $S^2 \times S^1$ topology (black rings) have been
constructed \cite{Emp}. It has been shown that for asymptotically
anti-de Sitter (AdS) spacetime, in the four-dimensional
Einstein-Maxwell theory, there exist black hole solutions whose
event horizons may have zero or negative constant curvature and
their topologies are no longer the two-sphere $S^2$. The
properties of these black holes are quite different from those of
black holes with usual spherical topology horizon, due to the
different topological structures of the event horizons. Besides,
the black hole thermodynamics is drastically affected by the
topology of the event horizon. It was argued that the Hawking-Page
phase transition \cite{Haw3} for the Schwarzschild-AdS black hole
does not occur for locally AdS black holes whose horizons have
vanishing or negative constant curvature, and they are thermally
stable \cite{Birm}. The studies on the topological black holes
have been carried out extensively in many aspects
\cite{Lemos,Cai2,Bril1,Cai3,Cai4,Cri,MHD,other,Ban}.

On the other hand, it is quite possible that gravity is not given
by the Einstein action, at least at sufficiently high energies. In
string theory, gravity becomes scalar-tensor in nature. The low
energy limit of the string theory leads to the Einstein gravity,
coupled non-minimally to a scalar dilaton field \cite{Wit1}. When
a dilaton is coupled to Einstein-Maxwell theory, it has profound
consequences for the black hole solutions. Many attempts to
construct exact solutions of Einstein-Maxwell-dilaton (EMd)
gravity have been made in the literature. For example, exact
solutions of EMd gravity in the absence of a dilaton potential
have been constructed in \cite{CDB1,CDB2}. The dilaton changes the
causal structure of the spacetime and leads to curvature
singularities at finite radii. These black holes are
asymptotically flat. In recent years, non-asymptotically flat
black hole spacetimes have received a lot of interest. There are
two motivations for exploring non asymptotically flat nor (A)dS
solutions of Einstein gravity. First, these solutions can shed
some light on the possible extensions of AdS/CFT correspondence.
Indeed, it has been speculated that the linear dilaton spacetimes,
which arise as near-horizon limits of dilatonic black holes, might
exhibit holography \cite{Ahar}. The second motivation comes from
the fact that such solutions may be used to extend the range of
validity of methods and tools originally developed for, and tested
in the case of, asymptotically flat or asymptotically AdS black
holes. Black hole spacetimes which are neither asymptotically flat
nor (A)dS have been explored by many authors
\cite{MW,PW,CHM,Cai,Clem,Mitra,Shey0,
SR,DF,Dehmag,SDR,SDRP,yaz,yaz2,SRM}. Thermodynamics of
$(n+1)$-dimensional black hole solutions with unusual asymptotics
have also been explored \cite{Shey,DHSR}.

In this paper, we would like to explore thermodynamics of the
topological dilaton black holes in higher dimensional spacetimes
in the presence of Liouville-type potentials for the dilaton
field. The motivation for studying higher dimensional solutions of
Einstein gravity originates from string theory, which is a
promising approach to quantum gravity. String theory predicts that
spacetime has more than four dimensions. For a while it was
thought that the extra spatial dimensions would be of the order of
the Planck scale, making a geometric description unreliable, but
it has recently been realized that there is a way to make the
extra dimensions relatively large and still be unobservable. This
is if we live on a three dimensional surface (brane) in a higher
dimensional spacetime (bulk) \cite{RS,DGP}. In such a scenario,
all gravitational objects such as black holes are higher
dimensional.

The outline of this paper is as follows: In Sec. \ref{Field}, we
construct a new class of $(n+1)$-dimensional topological black
hole solutions in EMd theory with two liouville type potentials
and general dilaton coupling constant, and investigate their
properties. In Sec. \ref{Therm}, we obtain the conserved and
thermodynamics quantities of the $(n+1)$-dimensional topological
black hole solutions and show that these quantities satisfy the
first law of thermodynamics. We also investigate the effect of the
dilaton field on the thermal stability of the solutions in this
section. The last section is devoted to summary and conclusions.

\section{Field Equations and Solutions\label{Field}}

The action of $(n+1)$-dimensional $(n\geq3)$
Einstein-Maxwell-dilaton gravity can be written \cite{CHM}
\begin{eqnarray}
S &=&\frac{1}{16\pi }\int d^{n+1}x\sqrt{-g}\left(
\mathcal{R}\text{ }-\frac{4}{n-1}(\nabla \Phi )^{2}-V(\Phi
)-e^{-4\alpha \Phi /(n-1)}F_{\mu \nu }F^{\mu \nu }\right) ,
\label{Act}
\end{eqnarray}
where $\mathcal{R}$ is the Ricci scalar curvature, $\Phi $ is the
dilaton field and $V(\Phi )$ is a potential for $\Phi $. $\alpha $
is a constant determining the strength of coupling of the scalar
and electromagnetic field, $F_{\mu \nu }=\partial _{\mu }A_{\nu
}-\partial _{\nu }A_{\mu }$ is the electromagnetic  field tensor
and $A_{\mu }$ is the electromagnetic potential. The equations of
motion can be obtained by varying the action (\ref{Act}) with
respect to the gravitational field $g_{\mu \nu }$, the dilaton
field $\Phi $ and the gauge field $A_{\mu }$ which yields the
following field equations
\begin{equation}
\mathcal{R}_{\mu \nu }=\frac{4}{n-1}\left( \partial _{\mu }\Phi
\partial _{\nu }\Phi +\frac{1}{4}g_{\mu \nu }V(\Phi )\right)
+2e^{-4\alpha \Phi /(n-1)}\left( F_{\mu \eta }F_{\nu }^{\text{
}\eta }-\frac{1}{2(n-1)}g_{\mu \nu }F_{\lambda \eta }F^{\lambda
\eta }\right) ,  \label{FE1}
\end{equation}
\begin{equation}
\nabla ^{2}\Phi =\frac{n-1}{8}\frac{\partial V}{\partial \Phi
}-\frac{\alpha }{2}e^{-{4\alpha \Phi }/({n-1})}F_{\lambda \eta
}F^{\lambda \eta }, \label{FE2}
\end{equation}
\begin{equation}
\nabla _{\mu }\left( e^{-{4\alpha \Phi }/({n-1})}F^{\mu \nu
}\right) =0.  \label{FE3}
\end{equation}
We would like to find topological solutions of the above field
equations. The most general such metric can be written in the form
\begin{equation}\label{metric}
ds^2=-f(r)dt^2 + {dr^2\over f(r)}+ r^2R^2(r)h_{ij}dx^{i}dx^{j} ,
\end{equation}
where $f(r)$ and $R(r)$ are functions of $r$ which should be
determined, and $h_{ij}$ is a function of coordinates $x_{i}$
which spanned an $(n-1)$-dimensional hypersurface with constant
scalar curvature $(n-1)(n-2)k$. Here $k$ is a constant and
characterizes the hypersurface. Without loss of generality, one
can take $k=0, 1, -1$, such that the black hole horizon or
cosmological horizon in (\ref{metric}) can be a zero (flat),
positive (elliptic) or negative (hyperbolic) constant curvature
hypersurface. The Maxwell equation (\ref{FE3}) can be integrated
immediately to give
\begin{eqnarray}
F_{tr} &=&\frac{q e^{4\alpha \Phi /(n-1)}}{(rR)^{n-1}},
\label{Ftr}
\end{eqnarray}
where $q$ is an integration constant related to the electric
charge of the black hole. Defining the electric charge via $ Q =
\frac{1}{4\pi} \int \exp\left[{-4\alpha\Phi/(n-1)}\right]  \text{
}^{*} F d{\Omega}, $ we get
\begin{equation}
{Q}=\frac{q\omega _{n-1}}{4\pi},  \label{Charge}
\end{equation}
where $w_{n-1}$ represents the volume of constant curvature
hypersurface described by $h_{ij}dx^idx^j$ . Our aim here is to
construct exact, $(n+1)$-dimensional topological solutions of the
EMd gravity with an arbitrary dilaton coupling parameter $\alpha$.
The case in which we find topological solutions of physically
interest is to take the dilaton potential of the form
\begin{equation}\label{v2}
V(\Phi) = 2\Lambda_{0} e^{2\zeta_{0}\Phi} +2 \Lambda e^{2\zeta
\Phi},
\end{equation}
where $\Lambda_{0}$,  $\Lambda$, $ \zeta_{0}$ and $ \zeta$ are
constants. This kind of  potential was previously investigated by
a number of authors both in the context of
Friedman-Robertson-Walker (FRW) scalar field cosmologies
\cite{ozer} and EMd black holes (see e.g
\cite{CHM,yaz2,SRM,Shey}). In order to solve the system of
equations (\ref{FE1}) and (\ref{FE2}) for three unknown functions
$f(r)$, $R(r)$ and $\Phi (r)$, we make the ansatz
\begin{equation}
R(r)=e^{2\alpha \Phi /(n-1)}.\label{Rphi}
\end{equation}
Using (\ref{Rphi}), the Maxwell fields (\ref{Ftr}) and the metric (\ref{metric}%
), one can easily show that equations (\ref{FE1}) and (\ref{FE2})
have solutions of the form
\begin{eqnarray}
f(r)&=&-{\frac { k(n-2)\left( { \alpha}^{2}+1 \right)
^{2}{b}^{-2\gamma}{r}^{2\gamma}}{\left( { \alpha}^{2}-1 \right)
\left({\alpha}^{2}+n-2 \right) }}-\frac{m}{r^{(n-1)(1-\gamma
)-1}}+\frac{2q^{2}(\alpha ^{2}+1)^{2}b^{-2(n-2)\gamma
}}{(n-1)(\alpha ^{2}+n-2)}r^{2(n-2)(\gamma -1)}\nonumber\\
&&+\frac{2\Lambda (\alpha ^{2}+1)^{2}b^{2\gamma }}{(n-1)(\alpha ^{2}-n)}%
r^{2(1-\gamma )}, \label{f}
\end{eqnarray}
\begin{equation}
\Phi (r)=\frac{(n-1)\alpha }{2(1+\alpha ^{2})}\ln (\frac{b}{r}),
\label{phi}
\end{equation}
where $b$ is an arbitrary constant and $\gamma =\alpha
^{2}/(\alpha ^{2}+1)$. In the above expression, $m$ appears as an
integration constant and is related to the ADM
(Arnowitt-Deser-Misner) mass of the black hole. According to the
definition of mass due to Abbott and Deser \cite{abot} (see also
\cite{Olea}), the mass of the solution (\ref{f}) is
\begin{equation}
{M}=\frac{b^{(n-1)\gamma}(n-1) \omega _{n-1}}{16\pi(\alpha^2+1)}m.
\label{Mass}
\end{equation}
In order to fully satisfy the system of equations, we must have
\begin{equation}\label{lam}
\zeta_{0} =\frac{2}{\alpha(n-1)},   \hspace{.8cm}
\zeta=\frac{2\alpha}{n-1}, \hspace{.8cm}    \Lambda_{0} =
\frac{k(n-1)(n-2)\alpha^2 }{2b^2(\alpha^2-1)}.
\end{equation}
Notice that here  $\Lambda$ is a free parameter which plays the
role of the cosmological constant. For later convenience, we
redefine it as $\Lambda=-n(n-1)/2l^2$, where $l$ is a constant
with dimension of length. One may note that in the absence of a
non-trivial dilaton ($\alpha=\gamma =0 $), the solution (\ref{f})
reduces to
\begin{eqnarray}
f(r)
&=&k-\frac{m}{r^{n-2}}+\frac{2q^2}{(n-1)(n-2)r^{2(n-2)}}-\frac{2\Lambda}{n(n-1)}r^2,
\end{eqnarray}
which describes an $(n+1)$-dimensional asymptotically AdS
topological black hole with a positive, zero or negative constant
curvature hypersurface (see for example \cite{Bril1,Cai3}).

\begin{figure}[tbp]
\epsfxsize=7cm \centerline{\epsffile{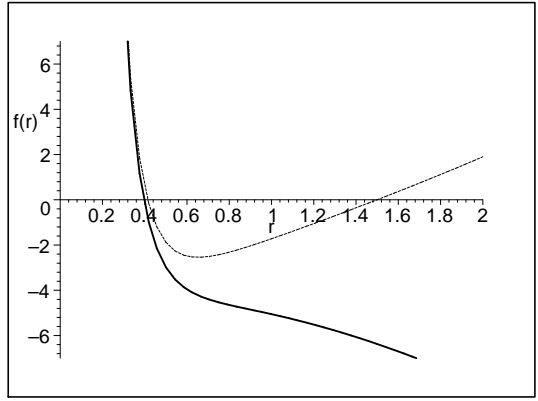}} \caption{The
function $f(r)$ versus $r$ for $k=-1$, $n=4$, $\protect\alpha=0.5$
and $m=2$. $\Lambda=+6$ (bold line), $\Lambda=-6$ (dashed line).}
\label{figureA}
\end{figure}

\begin{figure}[tbp]
\epsfxsize=7cm \centerline{\epsffile{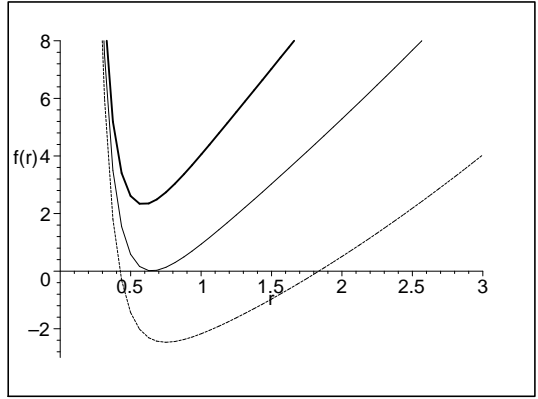}} \caption{The
function $f(r)$ versus $r$ for $\Lambda=-6$, $n=4$,
$\protect\alpha=0.67$ and $m=2$. $k=1$ (bold line), $k=0$
(continuous line) and $k=-1$(dashed line).} \label{figureB}
\end{figure}

\begin{figure}[tbp]
\epsfxsize=7cm \centerline{\epsffile{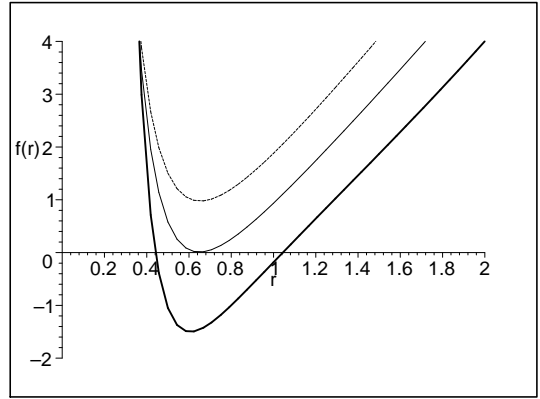}} \caption{The
function $f(r)$ versus $r$ for $\Lambda=-6$, $n=4$, $m=2$ and
$k=0$. $\protect\alpha=0.4$ (bold line), $\protect\alpha=0.67$
(continuous line) and $\protect\alpha=0.8$ (dashed line).}
\label{figureC}
\end{figure}

\begin{figure}[tbp]
\epsfxsize=7cm \centerline{\epsffile{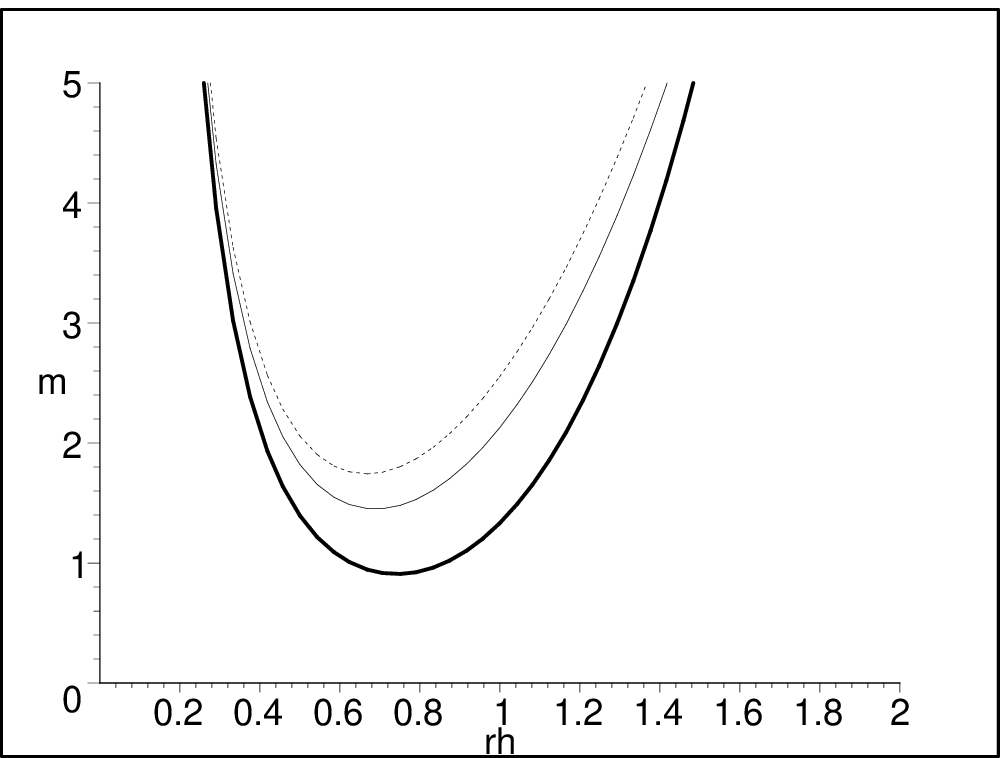}} \caption{The
function $m(r_h)$ versus $r_h$ for $\Lambda=-6$, $n=4$ and $k=0$.
$\protect\alpha=0$ (bold line), $\protect\alpha=0.5$ (continuous
line) and $\protect\alpha=0.6$ (dashed line).} \label{figureD}
\end{figure}

\begin{figure}[tbp]
\epsfxsize=7cm \centerline{\epsffile{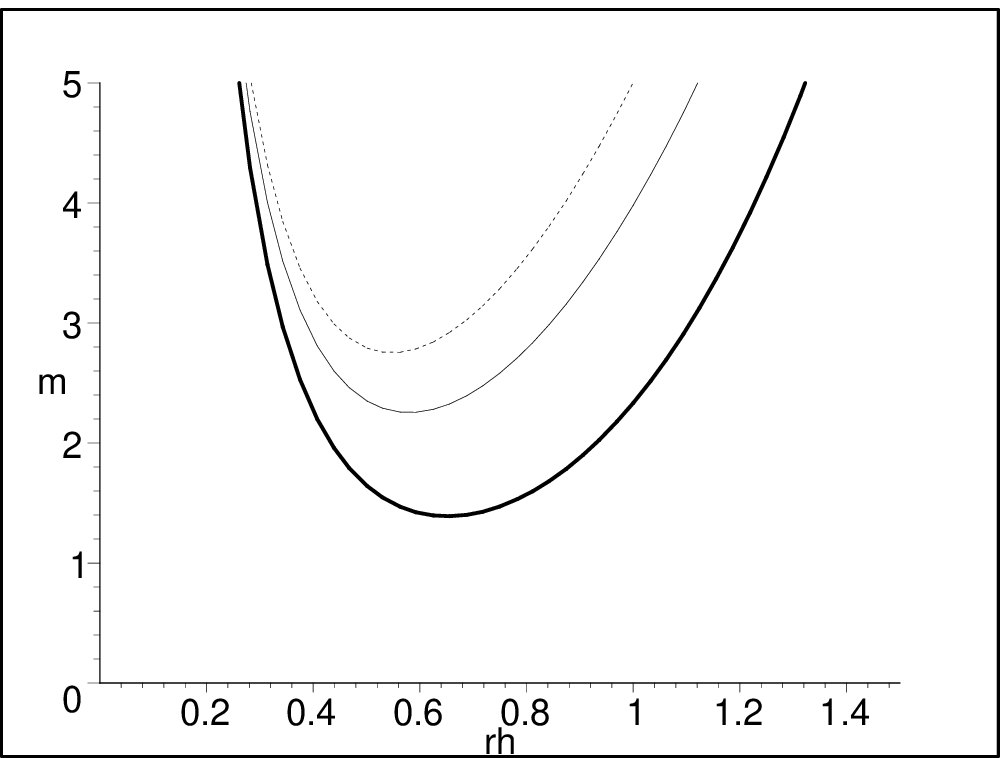}} \caption{The
function $m(r_h)$ versus $r_h$ for $\Lambda=-6$, $n=4$ and $k=1$.
$\protect\alpha=0$ (bold line), $\protect\alpha=0.5$ (continuous
line) and $\protect\alpha=0.6$ (dashed line).} \label{figureE}
\end{figure}

\begin{figure}[tbp]
\epsfxsize=7cm \centerline{\epsffile{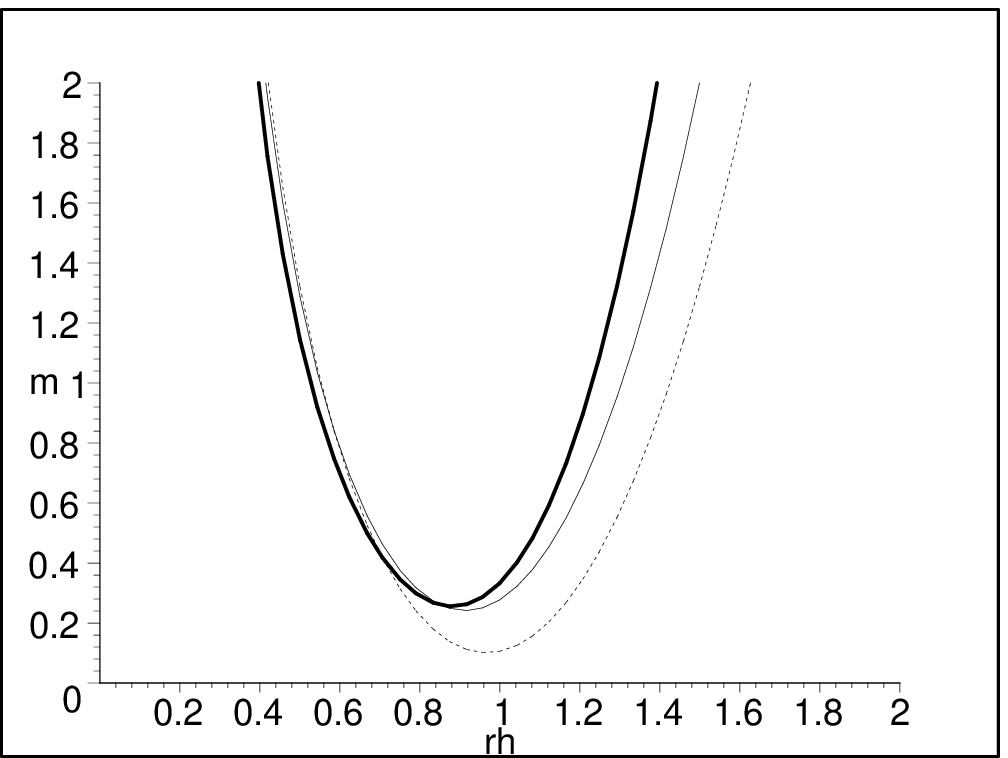}} \caption{The
function $m(r_h)$ versus $r_h$ for $\Lambda=-6$, $n=4$ and $k=-1$.
$\protect\alpha=0$ (bold line), $\protect\alpha=0.5$ (continuous
line) and $\protect\alpha=0.6$ (dashed line).} \label{figureF}
\end{figure}

\begin{figure}[tbp]
\epsfxsize=7cm \centerline{\epsffile{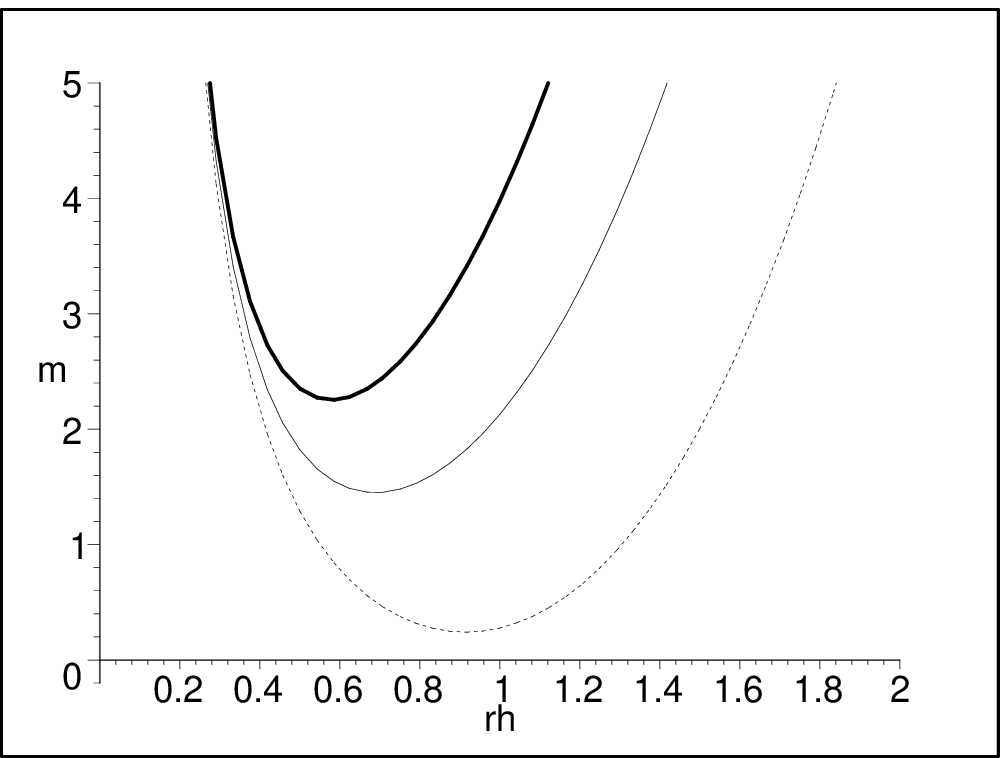}} \caption{The
function $m(r_h)$ versus $r_h$ for $\Lambda=-6$,
$\protect\alpha=0.5$ and $n=4$.  $k=1$ (bold line),   $k=0$
(continuous line) and
 $k=-1$ (dashed line).} \label{figureG}
\end{figure}

Next we study the physical properties of these solutions. To do
this, we first look for the curvature singularities. In the
presence of dilaton field, the Kretschmann scalar $R_{\mu \nu
\lambda \kappa }R^{\mu \nu \lambda \kappa }$ diverges at $r=0$, it
is finite for $r\neq 0$ and goes to zero as $r\rightarrow \infty
$. Thus, there is an essential singularity located at $r=0$. It is
notable to mention that in the $k=\pm1$ cases these solutions does
not exist for the string case where $\alpha=1$. As one can see
from Eq. (\ref{f}), the solution is ill-defined for $\alpha
=\sqrt{n}$. The cases with $\alpha <\sqrt{n}$ and $\alpha
>\sqrt{n}$ should be considered separately. In the first case
where $ \alpha <\sqrt{n}$,  there exist a cosmological horizon for
$\Lambda >0$, while there is no cosmological horizons if $\Lambda
<0$ (see fig. \ref{figureA}). Indeed, in the latter case ($\alpha
<\sqrt{n}$ and $\Lambda <0$) the spacetimes associated with the
solution (\ref{f}) exhibit a variety of possible casual structures
depending on the values of the metric parameters $\alpha $, $m$,
$q$ and $k$ (see figs. \ref{figureB}-\ref{figureC}). For
simplicity in these figures, we kept fixed the other parameters
$l=b=q=1$. These figures show that our solutions can represent
topological black hole, with inner and outer event horizons, an
extreme topological black hole or a naked singularity provided the
parameters of the solutions are chosen suitably. In the second
case where $\alpha
>\sqrt{n}$, the spacetime has a cosmological horizon for $\Lambda <0$ despite the value of
curvature constant $k$, while for $\Lambda>0$ we have cosmological
horizon in the case $k=1$ and naked singularity for $k=0,-1$. One
can obtain the casual structure by finding the roots of $ f(r)=0$.
Unfortunately, because of the nature of the exponent in (\ref{f}),
it is not possible to find analytically the location of the
horizons. To have further understanding on the nature of the
horizons, we plot in figures \ref{figureD}-\ref{figureG}, the mass
parameter $m$ as a function of the horizon radius for different
value of dilaton coupling constant $\alpha$ and curvature constant
$k$. Again, we have fixed $l=b=q=1$, for simplicity. It is easy to
show that the mass parameter $m$ of the topological black hole can
be expressed in terms of the horizon radius $r_{h}$ as
\begin{eqnarray}\label{mass}
m(r_{h}) &=&-{\frac { k(n-2)( \alpha^{2}+1)
^{2}{b}^{-2\gamma}}{\left( { \alpha}^{2}-1 \right)
\left(n+{\alpha}^{2}-2 \right)
}}{r_{h}}^{n-2+\gamma(3-n)}+\frac{2\Lambda \left( {\alpha}^{2}+1
\right) ^{2}{b}^{2 \gamma}}{(n-1)(\alpha^{2}-n
)}r_{h}^{n(1-\gamma)-\gamma} \nonumber\\&& +\frac{2q^2 \left(
{\alpha}^{2}+1 \right) ^{2}{b}^{-2 \gamma(n-2)}}{(n-1)
\left(n+{\alpha}^{2}-2 \right)}r_{h}^{(n-3)(\gamma-1)-1}.
\end{eqnarray}


These figures show that for a given value of $\alpha$, the number
of horizons depend on the choice of the value of the mass
parameter $m$. We see that, up to a certain value of the mass
parameter $m$, there are two horizons, and as we decrease the $m$
further, the two horizons meet. In this case we get extremal black
hole with mass $m_{\mathrm{ext}}$. These figures also show that
for $k=0,1$, with increasing $\alpha$, the $m_{\mathrm{ext}}$ also
increases, while for $k=-1$ it decreases as $\alpha$ increases.
Besides figure \ref{figureG} shows that for fixed value of the
other parameters, $m_{\mathrm{ext}}$ decreases with decreasing the
constant curvature $k$. Numerical calculations show that when we
have extremal topological black hole, the temperature of the black
hole vanishes. The Hawking temperature of the topological black
hole on the outer horizon $r_{+}$ can be calculated using the
relation
\begin{equation}
T_{+}=\frac{\kappa}{2\pi}= \frac{f^{\text{ }^{\prime
}}(r_{+})}{4\pi},
\end{equation}
where $\kappa$ is the surface gravity. Then, one can easily show
that
\begin{eqnarray}\label{Tem}
T_{+}&=&-\frac{(\alpha ^2+1)}{2\pi (n-1)}\left(
\frac{k(n-2)(n-1)b^{-2\gamma}}{2(\alpha^2-1)}r_{+}^{2\gamma-1}
+\Lambda b^{2\gamma}r_{+}^{1-2\gamma}+q^{2}b^{-2(n-2)\gamma
}r_{+}^{(2n-3)(\gamma -1)-\gamma}\right)\nonumber\\
&=&-\frac{k(n-2)(\alpha ^2+1)b^{-2\gamma}}{2\pi(\alpha
^2+n-2)}r_{+}^{2\gamma-1}+\frac{(n-\alpha ^{2})m}{4\pi(\alpha
^{2}+1)}{r_{+}}^{(n-1)(\gamma -1)} \nonumber\\
&&-\frac{q^{2}(\alpha ^{2}+1)b^{-2(n-2)\gamma }}{\pi(\alpha
^{2}+n-2)}{r_{+}}^{(2n-3)(\gamma -1)-\gamma}.
\end{eqnarray}
Equation (\ref{Tem}) shows that when $k=0$, the temperature is
negative for the two
cases of (\emph{i}) $\alpha >\sqrt{n}$ despite the sign of $\Lambda $, and (%
\emph{ii}) positive $\Lambda $ despite the value of $\alpha $. As
we argued above in these two cases we encounter with cosmological
horizons, and therefore the cosmological horizons have negative
temperature. Numerical calculations show that the temperature of
the event horizon goes to zero as the black hole approaches the
extreme case. It is a matter of calculation to show that
\begin{eqnarray}\label{mext}
m_{\mathrm{ext}}&=&\frac{2k(n-2)(\alpha
^2+1)^2b^{-2\gamma}}{(n-\alpha ^{2})(\alpha
^2+n-2)}r_{+}^{(2-n)(\gamma-1)+\gamma} +\frac{4q^2(\alpha
^{2}+1)^2b^{2(2-n)\gamma}}{(n-\alpha ^{2})(\alpha
^2+n-2)}r_{+}^{(3-n)(1-\gamma)-1}.
\end{eqnarray}
In summary, the metric of Eqs. (\ref{metric}) and (\ref{f}) can
represent a topological black hole with inner and outer event
horizons located at $r_{-}$ and $r_{+}$, provided
$m>m_{\mathrm{ext}}$, an extreme topological black hole in the
case of $m=m_{\mathrm{ext}}$, and a naked singularity if
$m<m_{\mathrm{ext}}$. It is worth noting that in the absence of a
non-trivial dilaton field ($\alpha=\gamma =0 $), expressions
(\ref{Tem}) and (\ref{mext}) reduce to that of an
$(n+1)$-dimensional asymptotically AdS topological black hole
\cite{Bril1,Cai3}.

\section{Thermodynamics of topological black hole} \label{Therm}
In this section we are going to explore thermodynamics of the
topological dilaton black hole we have just found. The entropy of
the topological black hole typically satisfies the so called area
law of the entropy which states that the entropy of the black hole
is a quarter of the event horizon area \cite{Beck}. This near
universal law applies to almost all kinds of black holes,
including dilaton black holes, in Einstein gravity \cite{hunt}. It
is a matter of calculation to show that the entropy of the
topological black hole is
\begin{equation}
{S}=\frac{b^{(n-1)\gamma}\omega _{n-1}r_{+}^{(n-1)(1-\gamma
)}}{4}.\label{Entropy}
\end{equation}
The electric potential $U$, measured at infinity with respect to
the horizon, is defined by
\begin{equation}
U=A_{\mu }\chi ^{\mu }\left| _{r\rightarrow \infty }-A_{\mu }\chi
^{\mu }\right| _{r=r_{+}},  \label{Pot}
\end{equation}
where $\chi=\partial_{t}$ is the null generator of the horizon.
One can easily show that the gauge potential $A_{t }$
corresponding to the electromagnetic field (\ref{Ftr}) can be
written as
\begin{eqnarray}\label{vectorpot}
A_{t}&=&\frac{qb^{(3-n)\gamma }}{\Upsilon r^{\Upsilon }},
\end{eqnarray}
where $\Upsilon =(n-3)(1-\gamma )+1$. Therefore, the electric
potential may be obtained as
\begin{equation}
U=\frac{qb^{(3-n)\gamma }}{ \Upsilon{r_{+}}^{\Upsilon }}.
\label{Pot}
\end{equation}
Then, we consider the first law of thermodynamics for the
topological black hole. In order to do this, we obtain the mass
$M$ as a function of extensive quantities $S$, and $Q$. Using the
expression for the charge, the mass and the entropy given in Eqs.
(\ref{Charge}), (\ref{Mass}) and (\ref{Entropy}) and the fact that
$f(r_{+})=0$, one can obtain a Smarr-type formula as
\begin{eqnarray}
M(S,Q)&=&-\frac{k(n-1)(n-2)(\alpha^2+1)b^{-\alpha^2}
}{16\pi(\alpha^2-1)(\alpha^2+n-2)}{\left(4S\right)}^{\frac{\alpha^2+n-2}{n-1}}
+\frac{\Lambda}{8\pi}\frac{(\alpha^2+1)
b^{\alpha^2}}{(\alpha^2-n)}{\left(4S\right)}^{\frac{n-\alpha^2}{n-1}}
\nonumber\\
&&+\frac{2\pi
Q^{2}(\alpha^2+1)b^{\alpha^2}}{\alpha^2+n-2}{\left(4S\right)}^{\frac{\alpha^2+n-2}{1-n}}.
\label{Msmar}
\end{eqnarray}
One may then regard the parameters $S$, and $Q$ as a complete set
of extensive parameters for the mass $M(S,Q)$ and define the
intensive parameters conjugate to $S$ and $Q$. These quantities
are the temperature and the electric potential
\begin{equation}
T=\left( \frac{\partial M}{\partial S}\right) _{Q},\ \ U=\left( \frac{\partial M%
}{\partial Q}\right) _{S}.  \label{Dsmar}
\end{equation}
Numerical calculations show that the intensive quantities
calculated by Eq. (\ref{Dsmar}) coincide with Eqs. (\ref{Tem}) and
(\ref{Pot}). Thus, these thermodynamics quantities satisfy the
first law of thermodynamics
\begin{equation}
dM = TdS+Ud{Q}.
\end{equation}
\begin{figure}[tbp]
\epsfxsize=7cm \centerline{\epsffile{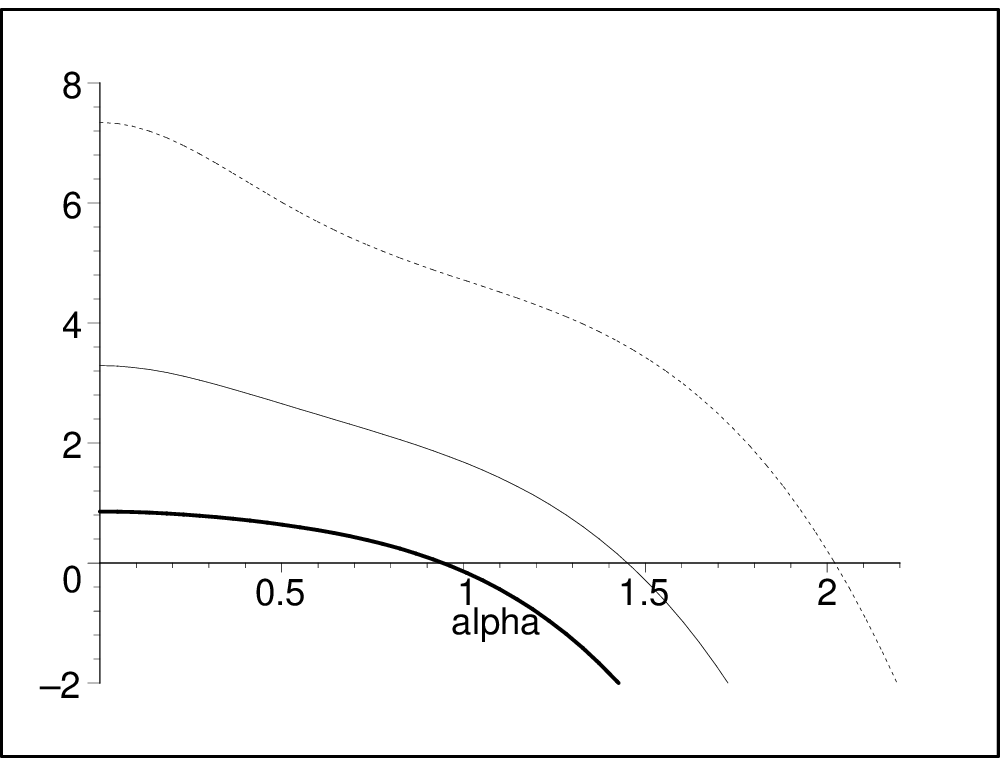}}
\caption{$(\partial ^{2}M/\partial S^{2})_{Q}$ versus
$\protect\alpha $ for $l=b=1$, $r_{+}=0.8$, $n=5$, and $k=1$.
$q=0.5$ (bold line), $q=1$ (continuous line), and $q=1.5$ (dashed
line).} \label{Figure1}
\end{figure}

\begin{figure}[tbp]
\epsfxsize=7cm \centerline{\epsffile{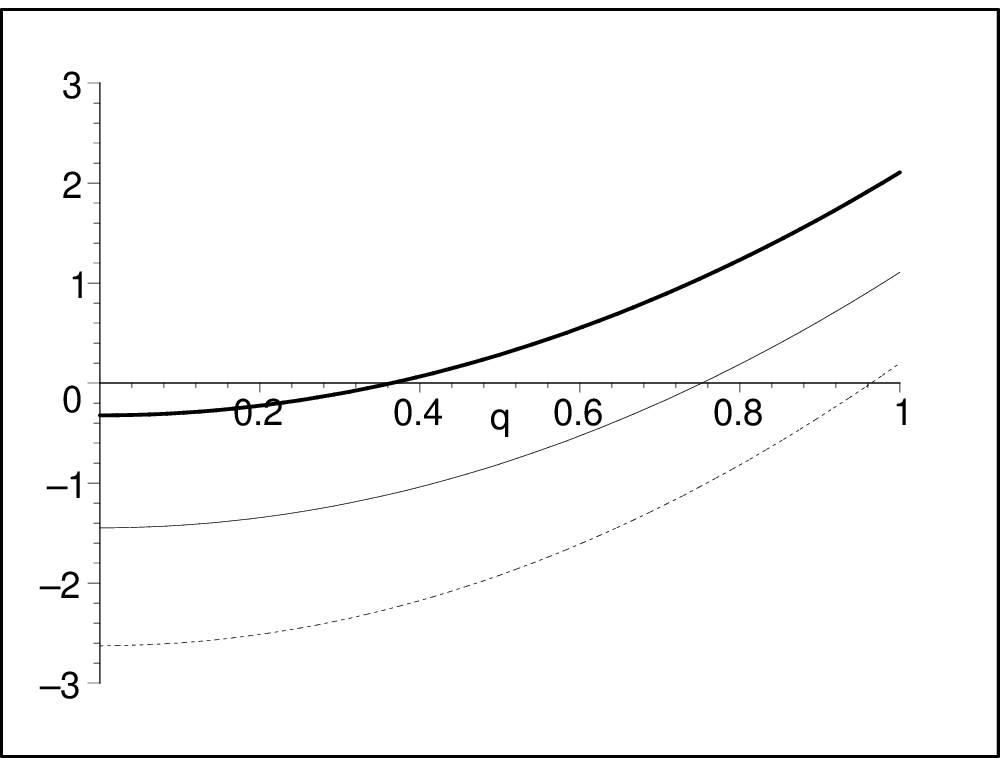}}
\caption{$(\partial ^{2}M/\partial S^{2})_{Q}$ versus $q$ for
$l=b=1$, $r_{+}=0.8$, $n=5$ and $k=1$. $\protect\alpha=0.8$ (bold
line), $\protect\alpha=1.2$ (continuous line) and
$\protect\alpha=\protect\sqrt{2}$ (dashed line).} \label{Figure2}
\end{figure}

\begin{figure}[tbp]
\epsfxsize=7cm \centerline{\epsffile{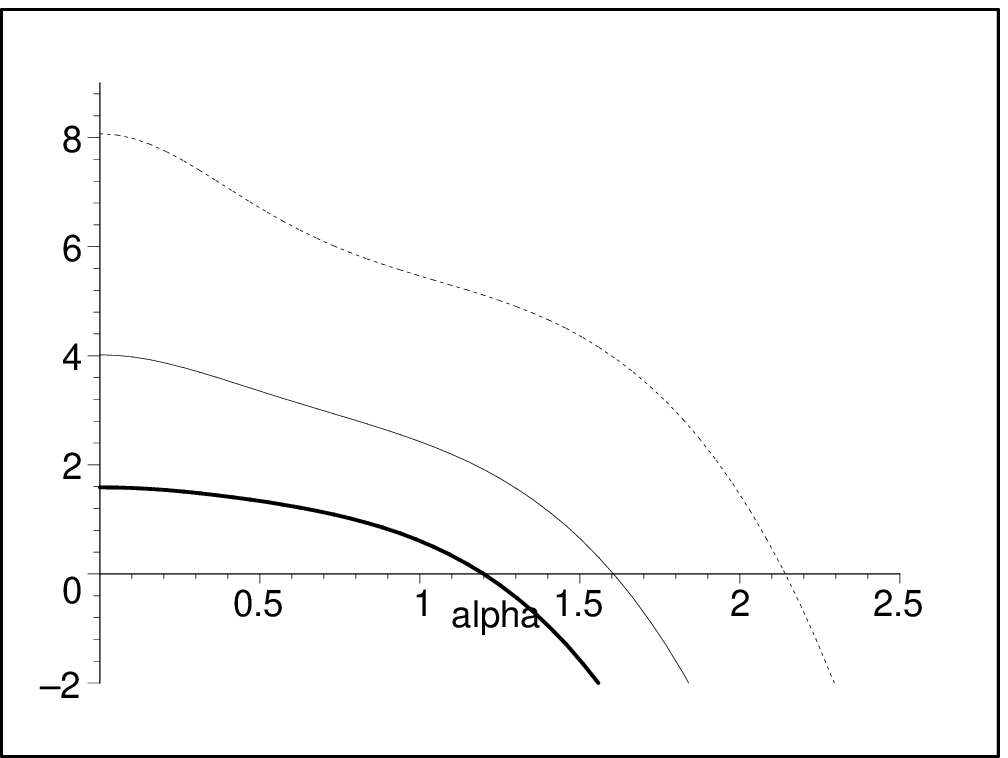}}
\caption{$(\partial ^{2}M/\partial S^{2})_{Q}$ versus
$\protect\alpha $ for $l=b=1$, $r_{+}=0.8$, $n=5$, and $k=0$.
$q=0.5$ (bold line), $q=1$ (continuous line), and $q=1.5$ (dashed
line).} \label{Figure3}
\end{figure}

\begin{figure}[tbp]
\epsfxsize=7cm \centerline{\epsffile{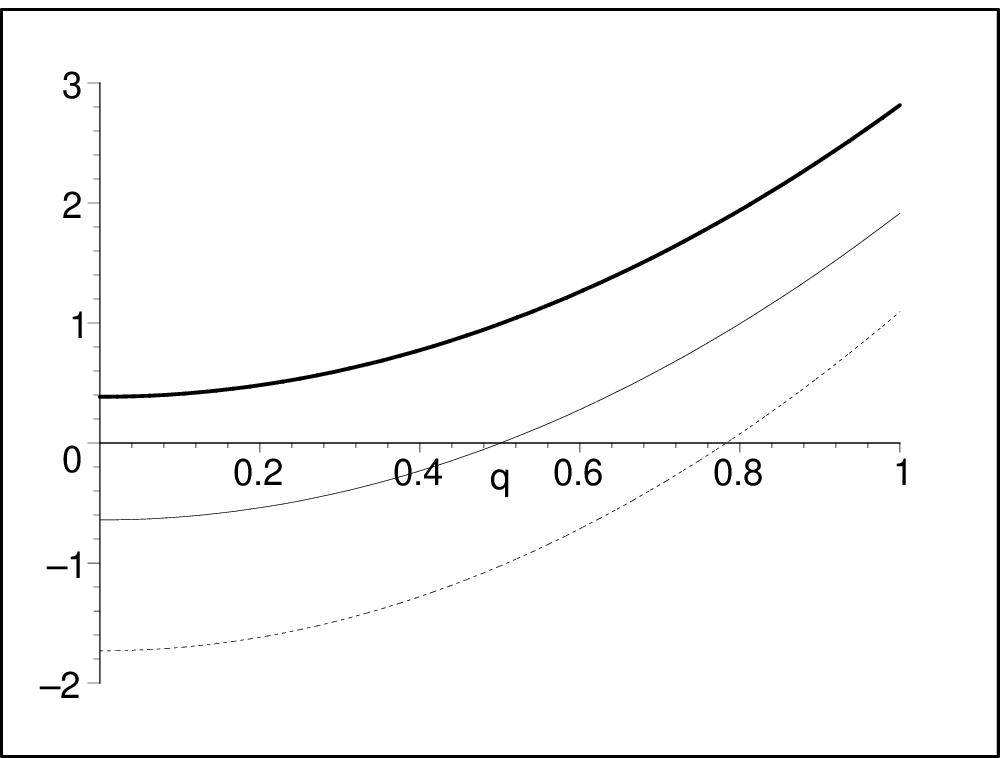}}
\caption{$(\partial ^{2}M/\partial S^{2})_{Q}$ versus $q $ for
$l=b=1$, $r_{+}=0.8$, $n=5$ and $k=0$. $\protect\alpha=0.8$ (bold
line), $\protect\alpha=1.2$ (continuous line) and
$\protect\alpha=\protect\sqrt{2}$ (dashed line).} \label{Figure4}
\end{figure}

\begin{figure}[tbp]
\epsfxsize=7cm \centerline{\epsffile{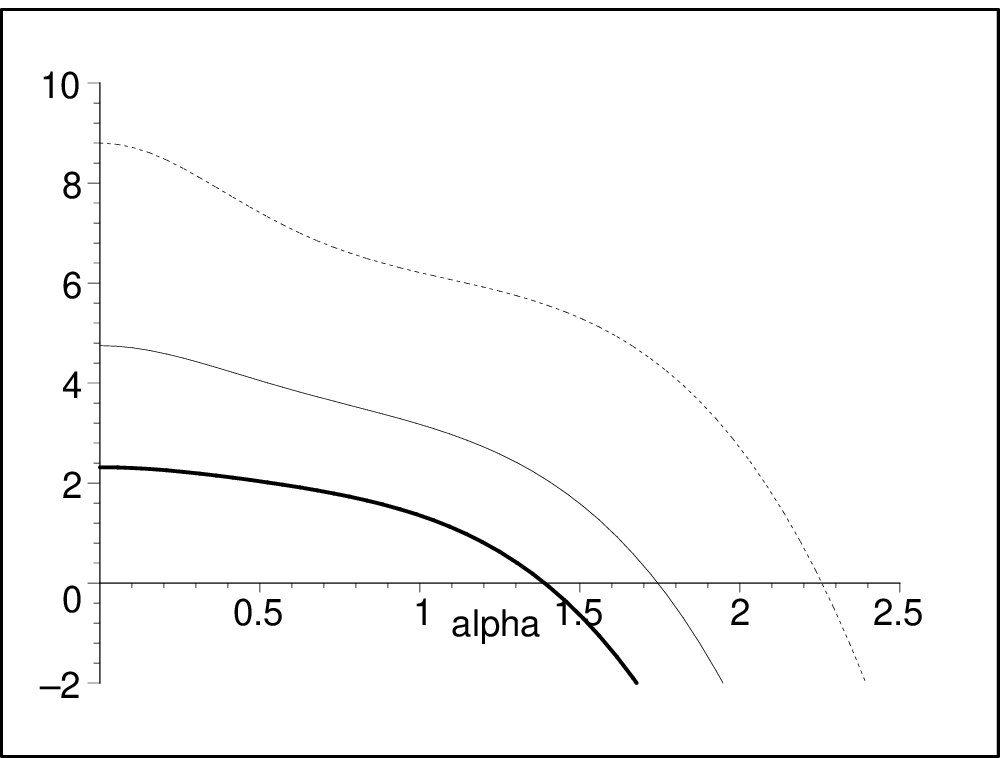}}
\caption{$(\partial ^{2}M/\partial S^{2})_{Q}$ versus
$\protect\alpha $ for $l=b=1$, $r_{+}=0.8$, $n=5$ and $k=-1$.
$q=0.5$ (bold line), $q=1$ (continuous line), and $q=1.5$ (dashed
line).} \label{Figure5}
\end{figure}

\begin{figure}[tbp]
\epsfxsize=7cm \centerline{\epsffile{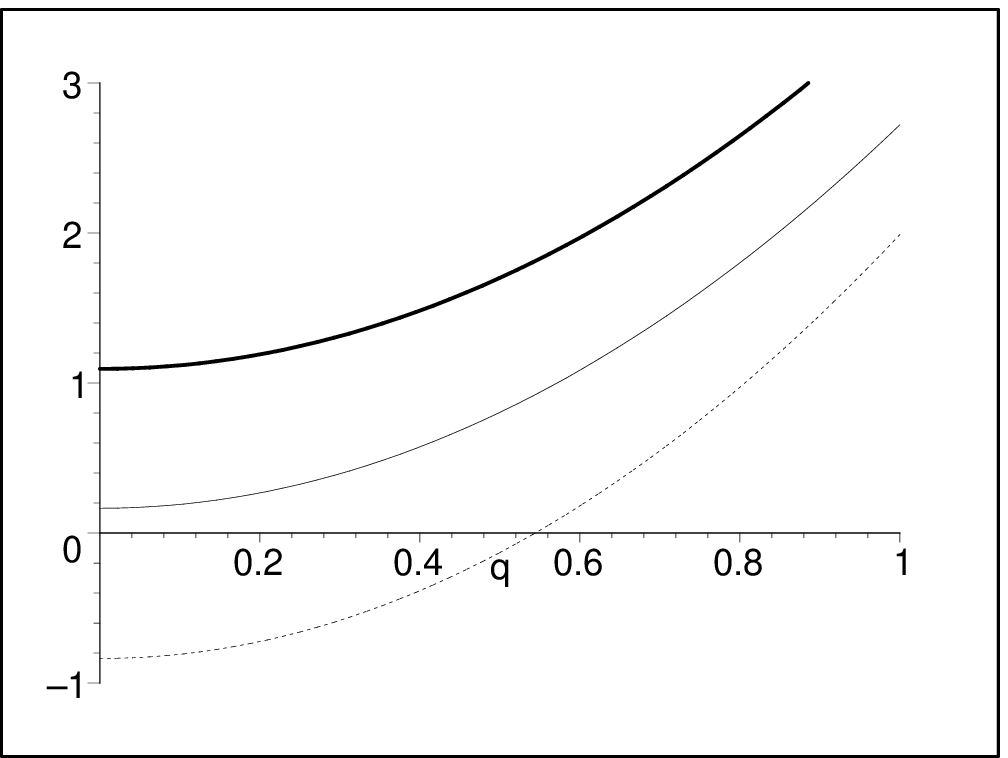}}
\caption{$(\partial ^{2}M/\partial S^{2})_{Q}$ versus $q $ for
$l=b=1$, $r_{+}=0.8$, $n=5$  and  $k =-1$. $\protect\alpha=0.8$
(bold line), $\protect\alpha=1.2$ (continuous line) and
$\protect\alpha=\protect\sqrt{2}$ (dashed line).} \label{Figure6}
\end{figure}

\begin{figure}[tbp]
\epsfxsize=7cm \centerline{\epsffile{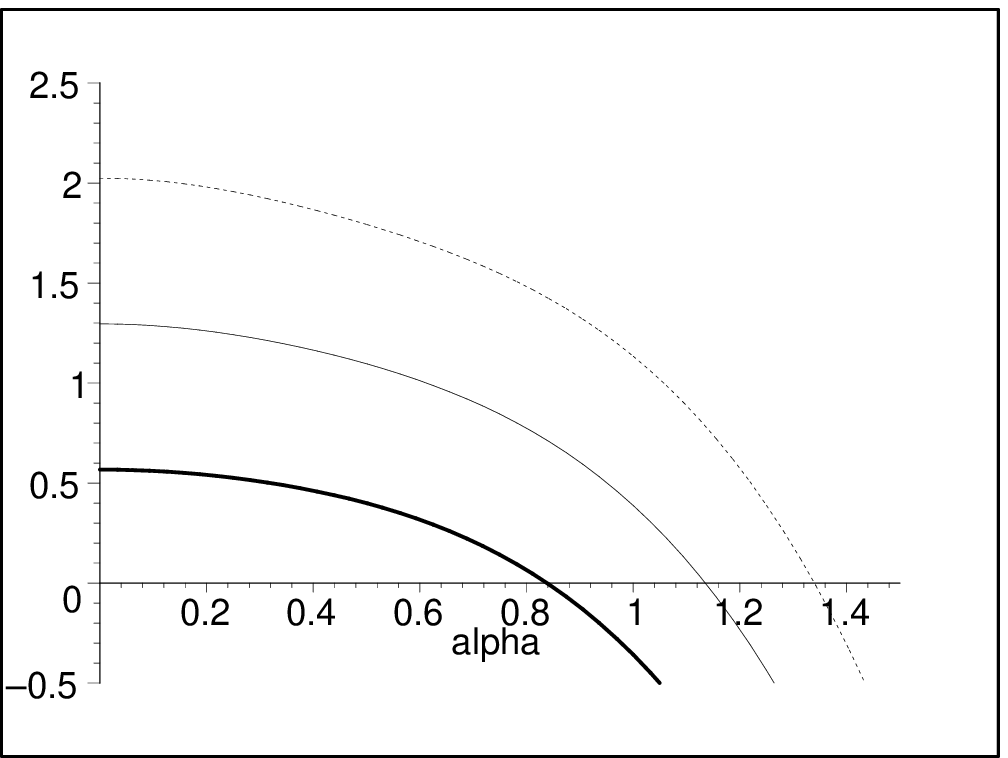}}
\caption{$(\partial ^{2}M/\partial S^{2})_{Q}$ versus $\alpha $
for $l=b=1$, $r_{+}=0.8$, $q=0.4$ and $n=5$. $k=1$ (bold line),
$k=0$ (continuous line), and $k=-1$ (dashed line).}
\label{Figure7}
\end{figure}

\begin{figure}[tbp]
\epsfxsize=7cm \centerline{\epsffile{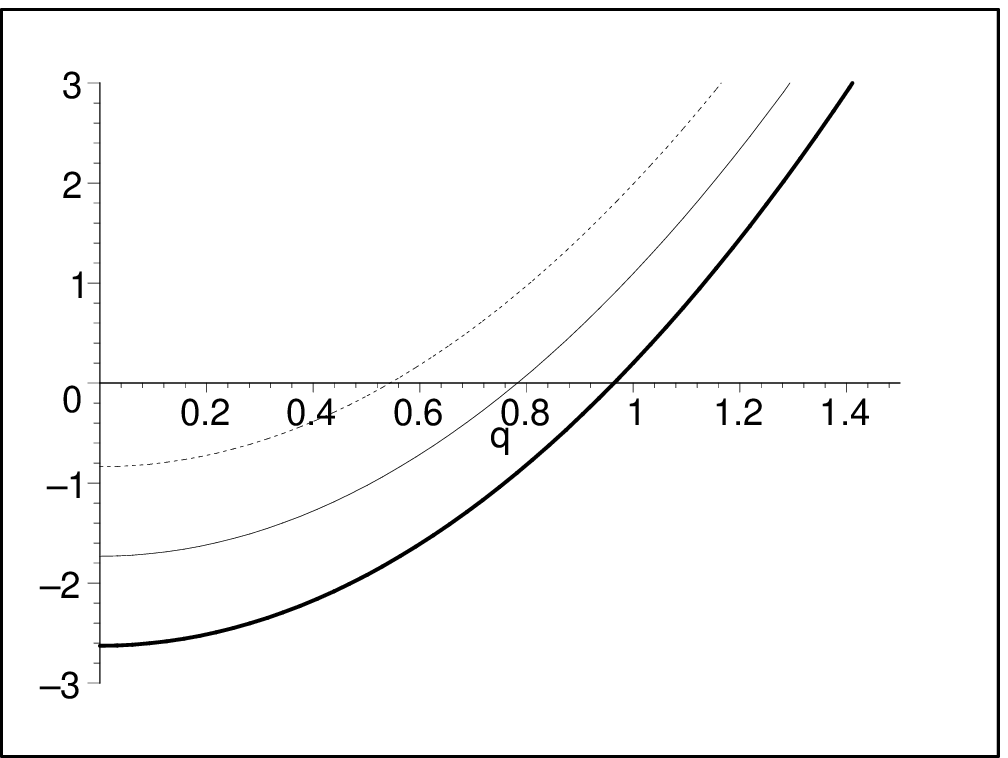}}
\caption{$(\partial ^{2}M/\partial S^{2})_{Q}$ versus $q $ for
$l=b=1$, $r_{+}=0.8$,  $\protect\alpha=\sqrt{2}$ and  $n=5$. $k=1$
(bold line), $k=0$ (continuous line), and $k=-1$ (dashed line).}
\label{Figure8}
\end{figure}
Finally, we study thermal stability of the topological dilaton
black hole. The stability of a thermodynamic system with respect
to small variations of the thermodynamic coordinates is usually
performed by analyzing the behavior of the entropy $ S(M,Q)$
around the equilibrium. The local stability in any ensemble
requires that $S(M,Q)$ be a convex function of the extensive
variables or its Legendre transformation must be a concave
function of the intensive variables. The stability can also be
studied by the behavior of the energy $M(S,Q)$ which should be a
convex function of its extensive variable. Thus, the local
stability can in principle be carried out by finding the
determinant of the Hessian matrix of $M(S,Q)$ with respect to its
extensive variables \cite{Cal2,Gub}. In our case the mass $M$ is a
function of entropy and charge. The number of thermodynamic
variables depends on the ensemble that is used. In the canonical
ensemble, the charge is a fixed parameter and therefore the
positivity of the $(\partial ^{2}M/\partial S^{2})_{Q}$ is
sufficient to ensure local stability. Numerical calculations show
that the topological black hole solutions are stable independent
of the value of the charge and curvature constant parameters $q$
and $k$ in any dimensions if $\alpha <\alpha_{\max }$, while for
$\alpha > \alpha_{\max }$ the system has an unstable phase. It is
notable to mention that for $k=0,-1$ we have $\alpha_{\max
}\geq1$. On the other hand, there is always a lower limit for the
electric charge, $q_{\min}$, for which the system is thermally
stable provided $q> q_{\min }$ (see figs.
\ref{Figure1}-\ref{Figure6}). It is worth noting that $\alpha
_{\mathrm{\max }}$ and $q_{\min}$  depend on the dimensionality of
the spacetime and the metric parameters $m$ and $l$. In figures
\ref{Figure7} and \ref{Figure8} we plot $(\partial ^{2}M/\partial
S^{2})_{Q}$ versus dilaton coupling constant $\alpha$ and charge
parameter $q$ for different value of the curvature constant $k$.
These figures show that for fixed value of the other parameters,
as we decrease the constant curvature $k$, the value of
$\alpha_{\max }$ increases while in contrast, $q_{\min}$
decreases.

\section{summary and conclusions}

In $(n+1)$-dimensional spacetime, when the $(n-1)$-sphere of black
hole event horizons is replaced by an $(n-1)$-dimensional
hypersurface with positive, zero or negative constant curvature,
the black hole is referred to as a topological black hole. In this
paper, first we obtained a new class of $(n+1)$-dimensional
$(n\geq3)$ topological black hole solutions in
Einstein-Maxwell-dilaton gravity with Liouville-type potentials
for the dilaton field. Then, we explored thermodynamics of these
topological dilaton black holes and disclosed the effect of the
dilaton field on the stability of the solutions. In contrast to
the topological black holes in the Einstein-Maxwell theory, which
are asymptotically AdS, the topological dilaton black holes we
found here, are neither asymptotically flat nor (A)dS. Indeed, the
Liouville-type potentials (the negative effective cosmological
constant) plays a crucial role in the existence of these black
hole solutions, as the negative cosmological constant does in the
Einstein-Maxwell theory. In the $k=\pm1$ cases, these solutions
does not exist for the string case where $\alpha=1$. In the
presence of Liouville-type potential, we obtained exact solutions
provided $\alpha \neq \sqrt{n}$. In the absence of a dilaton field
($\alpha =\gamma=0 $), our solutions reduce to the
$(n+1)$-dimensional topological black hole solutions presented in
\cite{Cai3}. We showed that our solutions can represent
topological black hole with inner and outer event horizons, an
extreme topological black hole or a naked singularity provided the
parameters of the solutions are chosen suitably. We also computed
the charge, mass, temperature, entropy and electric potential of
the topological dilaton black holes and found that these
quantities satisfy the first law of thermodynamics. We analyzed
the thermal stability of the solutions in the canonical ensemble
by finding a Smarr-type formula and considering $ (\partial
^{2}M/\partial S^{2})_{Q}$ for the charged topological dilaton
black hole solutions in $(n+1)$ dimensions. We showed that there
is no Hawking-Page phase transition in spite of charge of the
topological black hole provided $\alpha \leq \alpha_{\max }$,
while the solutions have an unstable phase for $\alpha>
\alpha_{\max }$. It is worth noting that for $k=0,-1$ we have
$\alpha_{\max }\geq1$. We found that there is always a low limit
for the electric charge, $q_{\min}$, for which the solutions are
stable provided $q>q_{\min }$.

\acknowledgments{This work was partially supported by Shahid
Bahonar University of Kerman.}

\end{document}